**Energy gap, penetration depth and surface resistance of MgB$_2$ thin films determined by microwave resonator measurements**

B.B. Jin[*] and N. Klein

Forschungszentrum Jülich, Institute of Thin Films and Interfaces, D-52425 Jülich, Germany,

W.N. Kang, Hyeong-Jin Kim, Eun-Mi Choi and Sung-IK Lee

National Creative Research Initiative Center for Superconductivity, Department of Physics, Pohang University of Science and Technology, Pohang, 790-784, Republic of Korea

* Correspondence author: b.b.jin @fz-juelich.de

*Abstract* - We have measured the temperature dependence of the microwave surface impedance $Z_s = R_s + i\omega\mu_0\lambda$ of two c-axis oriented MgB$_2$ films at a frequency $\omega/2\pi$ of 17.9 GHz employing a dielectric (sapphire) resonator technique. The temperature dependence of the magnetic field penetration depth $\lambda$ can be well fitted from 5 K close to $T_c$ by the standard BCS integral expression assuming the reduced energy gap $\Delta(0)/kT_c$ to be as low as 1.13 and 1.03 for the two samples. For the penetration depth at zero temperatures, values of 110 nm and 115 nm were determined from the fit. Our results clearly indicates the s-wave character of the order parameter. The temperature dependence of the surface resistance $R_s$ below $T_c/2$ is consistent with the low value of the energy gap. The surface resistance below 8 K was found to be below the resolution limit of 100 μΩ of our measurement technique.

The question about the energy gap in a particular superconducting material is of fundamental importance for the understanding of the relevant pairing mechanism and for the determination of its application potential. In the case of $MgB_2$ this question is raised with particular emphasis [1]. According to initial findings, $MgB_2$ seemed to comprise a high transition temperature with superconducting properties resembling that of conventional superconductors rather than that of high temperature superconducting cuprates [2-4]. In particular, the strong anisotropy of superconducting properties, the short coherence length and the unconventional order parameter being apparent for high temperature superconducting cuprates have turned out to be a burden for many applications. Consequently, such burdens are expected not to be present or at least less pronounced in $MgB_2$.

Recent experiments have brought some clarifications about the gap features of $MgB_2$. According to tunneling spectroscopy [5], point contact spectroscopy [6-7], photoemission [8], Raman scattering [9], specific heat [10-11], and optical conductivity spectra [12] there is evidence for an s-wave symmetry with two distinct gaps possibly associated with the two separate segments of the Fermi surface being present in $MgB_2$ [13]. The absolute values of these two gaps were determined to be around 1.5 to 3.8 meV and 5.5 to 8 meV consistently by these different methods.

On the other hand, the results for the temperature dependence of the penetration depth $\lambda(T)$ are quite controversial. Quadratic, linear and exponential dependence were reported in the literature [14-19]. Quadratic and linear dependence might be an indication for unconventional superconductivity, similar to high temperature superconducting cuprates [20]. However, the observation of an exponential dependence would be a clear indication for a nodeless gap, i.e. for an order parameter comprising s-wave symmetry. Hence, a very high measurement accuracy for temperature changes of $\lambda$ at $T \ll T_c$ applied to high quality samples is required to distinguish between a power-law and an exponential temperature dependence.

Microwave surface impedance measurements have proved to be the most sensitive tool to determine the temperature dependence of $\lambda$ of both thin film [21] and bulk single crystal samples [22]. In particular, they have been employed successfully to attain significant information about the symmetry of the order parameter in the high temperature superconducting cuprates. Therefore, microwave resonator techniques are most appropriate to be used for high-precision $\lambda(T)$ measurements on high quality $MgB_2$ samples.

Apart from the penetration depth the microwave surface resistance, $R_s$, is an important figure of merit for microwave applications. According to BCS theory, the expected $\exp(-\Delta/kT)$ dependence of $R_s$ below $T_c/2$ might result in very low $R_s$ values at temperature attainable with low-power cryocoolers.

$MgB_2$ thin films were fabricated using a two-step method by pulsed laser deposition. The detailed process is described in [23] and [24]. The films deposited on a plane parallel [1102] oriented sapphire substrate of 10x10 mm$^2$ in size exhibit a sharp resistive and inductive transition at transition temperatures up to 39 K (onset temperature of resistive transition). The film thickness was 400 nm for both samples. X-ray diffraction analysis indicates a high degree of c-axis-orientation perpendicular to the substrate surface and no detectable amount of MgO or any other crystalline impurity phases.

The microwave surface impedance was determined using a sapphire dielectric resonator technique described elsewhere [25]. The cavity with part of one endplate replaced by the thin film sample was excited in the $TE_{01\delta}$ - mode under weak coupling conditions. The unloaded quality factor $Q_0$ and resonant frequency was recorded as a function of temperature. The real part of the effective surface impedance, the effective surface resistance $R_s^{eff}$ was determined according to

$$R_s^{eff}(T) = G\left[\frac{1}{Q_0(T)} - \frac{1}{Q_{NbN}(4.2K)}\right] \qquad \text{Eq.1}$$

with $G$ being a geometrical factor determined by numerical simulation of the electromagnetic field distribution in the cavity and $Q_{NbN}$ (4.2 K) = 108,440 representing the unloaded quality factor measured by employing a high-quality NbN thin film as sample. The notation "effective" indicates an enhancement of $R_s$ and $\lambda$ due to the film thickness $d$ being of the order of $\lambda$ [26]. For temperatures below 30 K Eq. 1 allows for the determination of $R_s^{eff}$ with a systematic error of about 0.1 m$\Omega$, which is due to neglecting the temperature dependent background losses of the cavity and the small microwave losses ($R_s \approx 10^{-5} \Omega$) of the NbN film.

The temperature dependence of the effective penetration depth was determined from the temperature dependence of the resonant frequency $f(T)$ using

$$\delta\lambda^{eff}(T) = \frac{G}{\pi\mu_0}\frac{f(T)-f(5K)}{f^2(5K)} \qquad \text{Eq. 2}$$

with $\mu_0 = 1.256\cdot10^{-6}$ Vs/Am. There is a systematic error due to frequency changes caused by thermal expansion, the temperature dependence of the skin depth of the cavity wall material (copper) and the temperature dependence of the dielectric constant of sapphire. To account for these effects, we recorded the temperature dependence of the resonant frequency employing a copper sample. From this measurement we found that the systematic error is less than 1 nm for $T \leq 15$K and less than 2.5 nm for $T \leq 25$ K, which is at least one order of magnitude lower than the observed temperature changes of $\lambda$ for our MgB$_2$ samples. Therefore, Eq. 2 was applied without correction for our investigation.

In general, microwave resonator measurements do not allow for the determination of absolute values of $\lambda$ because the resonator dimensions are only know with a precision of several ten micrometers. However, absolute values of $\lambda$ can be extracted by comparing the measured temperature dependence with existing models.

Fig. 1 shows temperature dependence of $R_s^{eff}$ and $\delta\lambda^{eff}$ (sample B) as determined by Eq. 1 and Eq. 2, respectively. The inset shows the $R_s^{eff}(T)$ data at low temperatures. The full line represents the resolution limit of our setup. Below 8K, $R_s^{eff}$ is below the resolution of our technique (100$\mu\Omega$). Scaling to 10 GHz by assuming an $\omega^2$ law results in a corresponding value of less than 30µΩ. To our knowledge, this is the lowest value reported for wires, pellets and thin films of MgB$_2$ [17,27-29].

Fig.2 shows the measured temperature dependence of $\delta\lambda^{eff}$ determined from $f(T)$ according to Eq. 2. in a log $\delta\lambda$ versus $1/T$ representation. The full lines represent BCS calculations based on the standard BCS integral expression [20]

$$\frac{1}{\lambda^2(T)} = \frac{1}{\lambda^2(0)}\left[1 - 2\int_{\Delta(T)}^{\infty} -\frac{\partial f(\varepsilon)}{\partial \varepsilon}\frac{\varepsilon}{\sqrt{\varepsilon^2 - \Delta^2(T)}}d\varepsilon\right] \quad \text{Eq.3}$$

with $f(\varepsilon)=[\exp(\varepsilon/kT)+1]^{-1}$ representing the Fermi function and $\Delta(T)$ the temperature dependence of energy gap. The quantities $\Delta(0)/kT_c$ and $\lambda(0)$ are used as fit parameters. For the temperature dependence of $\Delta(T)/\Delta(0)$ values tabulated in Ref [30] were used. The $\delta\lambda_{BCS}$ values calculated from Eq. 3 are multiplied with $\coth(d/\lambda_{BCS})$ to account for the finite film thickness [26]. As a result, the fits of Eq. 3 to the $\delta\lambda_{eff}(T)$ data (full lines in Fig. 2) yield $\Delta(0)/kT_c$=1.13, $T_c$=39K ($\Delta(0)$=3.8meV), and $\lambda(0)$=110nm for sample A, $\Delta(0)/kT_c$=1.03, $T_c$=36K ($\Delta(0)$=3.2meV) and $\lambda(0)$=115nm for sample B. According to Fig. 2., there is a very good agreement between experimental data and the fit curves over the entire temperature range. The downturn of the experimental data very close to $T_c$ can be explained by impedance transformations resulting from the finite film thickness [26]. The nearly linear dependence below $T_c/2$ in this representation clearly indicates the s-wave nature of the order parameter, i.e. a finite energy gap in all directions of the Fermi surface without nodes. In contrast, a double logarithmic plot of the same data indicates that a power law does not fit the data at low temperatures. For comparison, the same analysis was performed for a NbN thin film, which is a well-known s-wave superconductor with a relatively high $T_c$ of 15.8 K. The fit parameters are $\Delta(0)/kT_c = 2.3$ and $\lambda(0) = 265$ nm, respectively, which is similar to literature data [31].

The $\Delta(0)/kT_c$ values obtained for the MgB$_2$ samples are similar to the value of 3.5±0.5meV determined by scanning tunneling microscopy [5]. The $\lambda(0)$ values are also in the range of corresponding values determined by other techniques [14,17,32,33]. In contrast to tunneling experiments, which are very sensitive to a possible modification of the surface, microwave surface impedance measurement probe almost the entire volume of a thin film sample with thickness of the order of $\lambda$. Therefore, surface degradation can be ruled out as a possible reason for the small values of $\Delta/kT_c$. In fact, some tunneling data exhibit $\Delta/kT_c$ values even lower than ours, indicating that surface degradation effects may play some role there [3,4].

A possible explanation for $\Delta(0)/kT_c$ being much smaller than the BCS value of 1.76 is that the small gap represents the smallest component of a double-or multigap or a strongly anisotropic gap. In this case the temperature dependence of $\lambda$ at $T<<T_c$ would be determined by its minimum value, because $\lambda(T)$ probes the thermal excitations with the lowest activation energy. However, the existence of a second larger gap should have a significant impact on $\lambda(T)$ at higher temperatures. Fig.3 shows the temperature dependence of $\lambda^2(0)/\lambda^2(T)$, which represents the normalised superfluid density. As expected from Fig. 2, a very good agreement with a single gap BCS theory is achieved for both samples. Similar to the $MgB_2$ samples, the NbN film also exhibits a very good agreement over the entire temperature range. The latter indicates that the effects of strong coupling on $\lambda(T)$ can be still modelled by the BCS integral expression just by assuming $\Delta/kT_c$ being larger than the BCS-value of 1.76 [34].

For comparison, the power law dependence $[\lambda(0)/\lambda(T)]^2 = 1- (T / T_c)^n$ with $n = 2$ ("quadratic") and $n = 4$ ("two fluid model") is depicted in Fig. 3. Above about $T_c/2$, $n = 4$ represents a fairly good approximation for the case of strong coupling, as observed for NbN. As quoted in [19], $n = 2$ can be used to fit the $MgB_2$ data above about $T_c / 2$ due to an appropriate choice of $\lambda(0)$. However, there is a clear discrepancy due to the exponential dependence below $T_c/ 2$ being apparent in our experimental data, ruling out the possibility of unconventional superconductivity in $MgB_2$.

Obviously, our data do not show any indication of a second larger energy gap. However, since the films have a high degree of c-axis orientation and since the rf current induced by the resonator field are "in-plane", we cannot exclude a strongly anisotropic gap. In the framework of possible two-gap superconductivity in $MgB_2$ thermal excitations from the small energy gap may "smear" the effects of the large energy gap in the quasiparticle density of states.

The absence of a second energy gap was also found for SIS junction [7,35]. It is believed that the persistent and clear observation of only one small energy gap is due to the much higher probability of tunneling into the 3d band, where only the small energy gap exists [7].

Fig. 4 shows temperature dependence of the $R_s^{eff} - R_s^{eff}$ (4.2 K) date of our samples. The full lines indicates fits of an exp($-\Delta/kT$) dependence with $\Delta/kT_c$ values as determined from $\lambda(T)$. According to Fig. 4, the data are consistent with the fit. However, the larger scattering of data does not allow for a precise determination of $\Delta/kT_c$ from $R_s^{eff}(T)$ on its own.

In conclusion, our experimental finding represents clear evidence for the existence of a finite gap with $\Delta(0)/kT_c$ values around 1.1. The temperature dependence of $\lambda$ can be well fitted by BCS theory in the entire temperature range below $T_c$. The very low $R_s$ value (<100μΩ at 4.2K and 270μΩ at 15K and 17.9GHz) obtained for one sample implies that $MgB_2$ is a promising material for microwave applications.

**Acknowledgement**

We thank for Dr. Z.Wang, KARC communications Research lab, Japan, for providing NbN thin films. This work at Postech was supported by the Ministry of Science and Technology of Korea through the Creative Research Initiative Program.

**Figure captions:**

Fig. 1: Temperature dependence of $R_s^{eff}$ and $\delta\lambda^{eff}$ of a MgB$_2$ thin film recorded at 17.9 GHz using a dielectric resonator technique. The inset shows the $R_s^{eff}$ (T) values below 8K and our measurement resolution of 100 µΩ.

Fig. 2: log $\delta\lambda^{eff}$ vs. 1/$T$ representation of the penetration depth data for two MgB$_2$ samples and a NbN thin film for comparison. The solid line represents BCS fits with parameters as described in the text.

Fig. 3: $\lambda^2(0)/\lambda^2(T)$ dependence for the samples shown in Fig. 2. The full lines represent BCS fits using $\Delta(0)/kT_c$ as fit parameter. The predictions by two fluid model (dashed), quadratic (dashed-dotted) dependence and standard BCS model with $\Delta(0)/kT_c$ = 1.76 (dotted) are also drawn.

Fig.4: $R_s^{eff}$ (T) - $R_s^{eff}$ (T = 4.2 K) plotted versus $T_c/T$ for the samples shown in Fig. 2. The full lines correspond to fits of exp(-$\Delta$(0) /$kT_c$) to the experimental data.

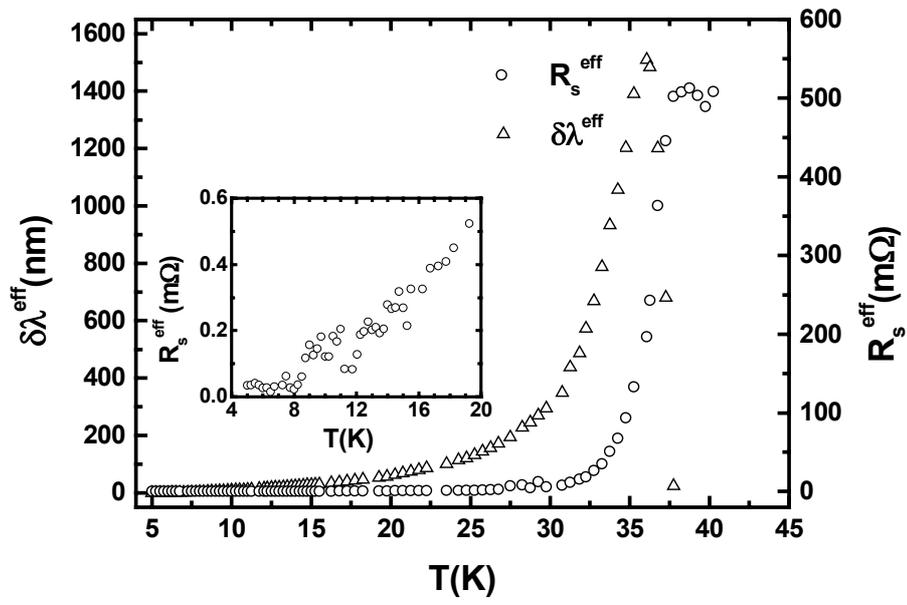

Fig.1

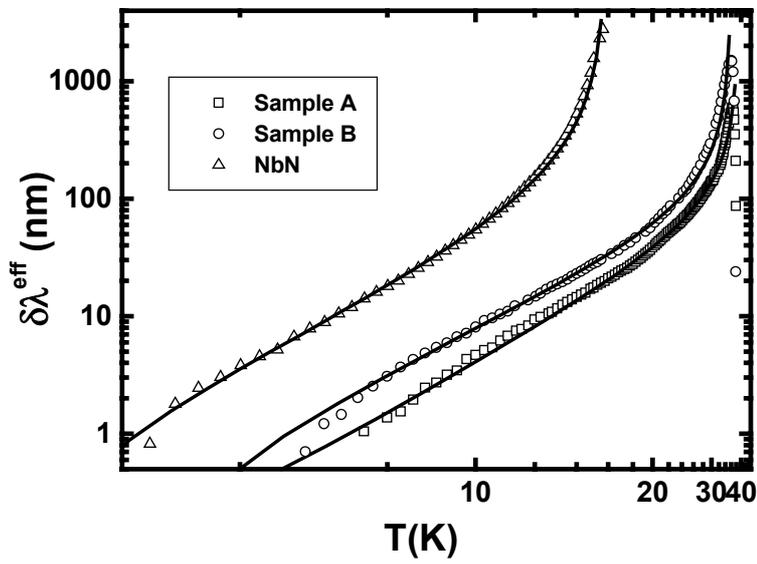

Fig.2

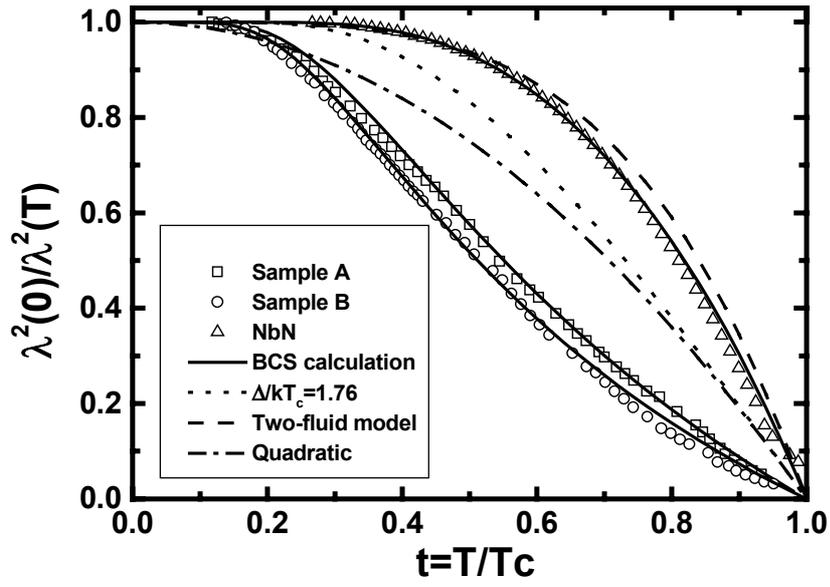

**Fig.3**

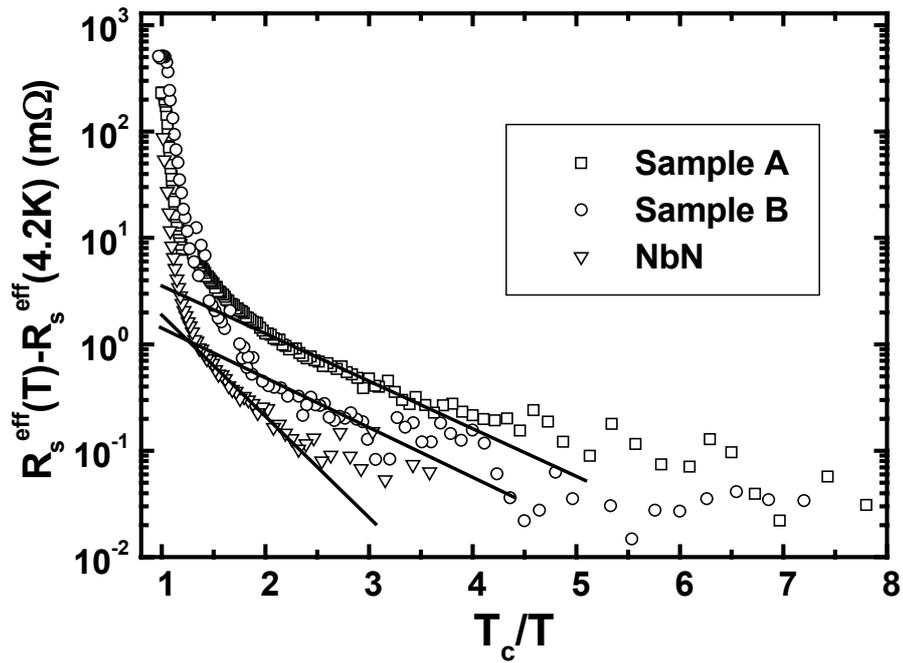

**Fig.4**